\documentclass[twocolumn, pra, showpacs,superscriptaddress]{revtex4}

\usepackage{amsmath}
\usepackage{amssymb}
\usepackage{graphicx}

\begin{document}

\title{Interspecies singlet pairing in a mixture of two spin-1 Bose condensates}
\author{Jie Zhang}
\affiliation{Institute of Theoretical Physics, Shanxi University, Taiyuan 030006,
People's Republic of China}
\author{Tiantian Li}
\affiliation{Institute of Theoretical Physics, Shanxi University, Taiyuan 030006,
People's Republic of China}
\author{Yunbo Zhang}
\email{ybzhang@sxu.edu.cn}
\affiliation{Institute of Theoretical Physics, Shanxi University, Taiyuan 030006,
People's Republic of China}

\begin{abstract}
We study the ground state properties of a mixture formed by two spin-1
condensates in the absence of an external magnetic field. As the collisional 
symmetry between interspecies bosonic atoms is broken, the interspecies coupling
interaction ($\beta $) and interspecies singlet pairing interaction ($\gamma
$) arise. The ground state can be calculated using the angular momentum theory
analytically for $\gamma =0$. The full quantum approach of exact diagonalization is adopted
numerically to consider the more general case as $\gamma \neq 0$. We
illustrate the competition between the two interspecies interactions and
find that as singlet pairing interaction dominates (or the total spin
vanishes), there are still different types of singlet formations
which are well determined by $\beta $.
\end{abstract}

\maketitle

\section{Introduction}

Since the MIT group succeeded in trapping a $^{23}$Na condensate in an
optical potential \cite{Stamper-Kurn1}, the spin degrees of freedom are
liberated, that give rise to a rich variety phenomena \cite{UedaRev} such as spin domains
\cite{Stamper-Kurn2}, textures \cite{Ohmi}, spin mixing dynamics \cite%
{Law98,Pu99,WenxianZhang}, and fragmentation of condensate \cite{HoYip,Mueller06}. 
The properties of such a three-component spinor
condensate were first studied using mean-field theories (MFT) \cite{Ho,Ohmi}
and polar and ferromagnetic spinor condensates had been implemented
experimentally \cite{Stenger,MSChang}. It was initially predicted that the
ground state of $^{23}$Na BEC ($c_{2}>0$) is either polar ($n_{0}=N$) or
anti-ferromagnetic ($n_{1}=n_{-1}=N/2)$ in the mean-field theory. However,
the results from many body theory \cite{Law98} pointed out that the
ground state of $^{23}$Na atoms is a spin singlet with properties
drastically different from those of mean field theories ($
n_{1}=n_{0}=n_{-1}=N/3 $). It was shown \cite{HoYip} that the singlet
ground state in zero field is a fragmented condensate with anomalously large
number fluctuations ($\Delta n_{\alpha }\sim N$) and thus has fragile
stability. The exact quantum eigenstates in the spin-2 case \cite{Koashi,Ueda} 
are also found and their magnetic response to a weak
magnetic field is compared with their mean field counterpart \cite{Ciobanu}.

Mixtures of scalar condensates with more than one atomic species or state
are actively studied theoretically \cite{Ho96,Pu2,Esry,Timmermans}.
Experimentally, the Feshbach resonance has exploited to create a double
species condensate with tunable interactions and the dynamics of the
superfluid and controllable phase separation are observed \cite%
{Myatt,Modugno,Thalhammer,Papp}. By adjusting the two
s-wave scattering lengths $a_{0}$ and $a_{2}$ through the so-called optical
Feshbach resonances \cite{OFR}, the spin exchange interaction between
individual atoms can be precisely tuned. The theoretical studies 
on mixtures of spinor condensates attracted much attention recently and both mean field and 
quantum many body theories have been applied to this novel system \cite{Luo,Xu09,XuZF,XuBA,zj}.
A temporal modulation of spin 
exchange interaction, which is tunable with optical Feshbach resonance, was 
recently proposed to localize the spin mixing dynamics in a $^{87}$Rb condensate \cite{Zhang2010}.

The interspecies scattering parameters between $^{87}$Rb and $^{23}$Na are
calculated resorting to the simple approach of the degenerate internal-state
approximation (DIA) \cite{DIA,DIA1,DIA2} as the low-energy atomic
interactions can be mostly attributed to the ground-state configurations of
the two valence electrons. The interspecies scattering lengths for singlet
and triplet electronic states are approximately determined already \cite%
{DIA1,DIA2}, given by $a_{S}=109a_{0}$ and $a_{T}=70a_{0}$, where $a_{0}$ is
the Bohr radius. The interspecies interactions between $^{87}$Rb and $^{23}$%
Na atoms are then parametrized by three scattering lengths (see eq. (\ref%
{scatter}) bellow), each being a linear combination of $a_{S}$ and $a_{T}$
weighted by the appropriate $9j$ coefficients for the total combined spins
of $F=0, 1$, and $2$. Within this approximation, it is found
coincidently that, the parameter for the interspecies singlet-pairing
interaction $\gamma $, is equal to zero \cite{Luo}.

However, for the more general case of an arbitrary mixture of spin-1
condensate, the hyperfine interaction between nuclear spin and electron spin
gives in general non-neglegible interspecies pairing and DIA approximation
is not applicable \cite{DIA,DIA1,DIA2,Luo}. In this paper, we first study
the various quantum phases of the binary mixture of spin-1 condensates in
the ground state ignoring the interspecies singlet pairing. Then the
situation with competition between intra- and inter-species singlet pairings
is considered. Using the full quantum approach of exact diagonalization, we
present the detailed phases diagram for the $\gamma \neq 0$ case.

\section{The Hamiltonian}

We take the intra-condensate atomic interaction in the form $V_{j}(\mathbf{r}%
)=(\alpha _{j}+\beta _{j}\mathbf{F}_{j}\cdot \mathbf{F}_{j})\delta (\mathbf{r%
})$ with $j=1,2$ for the two species and the inter-species interaction is
described as
\begin{equation}
V_{12}(\mathbf{r})=\frac{1}{2}%
(g_{0}^{(12)}P_{0}+g_{1}^{(12)}P_{1}+g_{2}^{(12)}P_{2})\delta (\mathbf{r}).
\label{scatter}
\end{equation}%
In contrast to intra-condensate interactions between identical atoms \cite%
{Luo,Xu09}, the collision between atoms belonging to different species can
occur in the total spin $F=1$ channel, which makes the
mixture more interesting. Here%
\begin{equation}
g_{0,1,2}^{(12)}=4\pi \hbar ^{2}a_{0,1,2}^{(12)}/\mu ,  \label{a}
\end{equation}
with $a_{0,1,2}^{(12)}$ is the scattering lengths in the channels of total
spin $F =0, 1$, and $2$, respectively, and $\mu
=M_{1}M_{2}/(M_{1}+M_{2}) $ denotes the reduced mass for the pair of atoms,
one each from the two different species with masses $M_{1}$ and $M_{2}$
respectively. $P_{0,1,2}$ is the corresponding projection operator with the
relationship $1=P_{2}+P_{1}+P_{0}$ and $\mathbf{F}_{1}\cdot \mathbf{F}%
_{2}=P_{2}-P_{1}-2P_{0}$, from which we get
\begin{equation}
V_{12}(\mathbf{r})=\frac{1}{2}(\alpha +\beta \mathbf{F}_{1}\cdot \mathbf{F}%
_{2}+\gamma P_{0})\delta (\mathbf{r})
\end{equation}%
with the parameters $\alpha =(g_{1}^{(12)}+g_{2}^{(12)}$ $)/2,\beta
=(-g_{1}^{(12)}+g_{2}^{(12)}$ $)/2$ and $\gamma $ $%
=(2g_{0}^{(12)}-3g_{1}^{(12)}+g_{2}^{(12)}$ $)/2$. $P_{0}$ projects an
inter-species pair into spin singlet state \cite{Xu09}. Furthermore, these parameters 
can be related to the singlet and triplet scattering lengths by means of a method 
based on $9j$ coefficient \cite{Luo}
\begin{eqnarray}
\alpha &=&\frac{\pi \hbar ^{2}}{\mu }%
(3a_{T}+a_{S})  \notag \\
\beta &=&\frac{\pi \hbar ^{2}}{4\mu }%
(a_{T}-a_{S})  \notag \\
\gamma &=&0
\end{eqnarray}

Denote the atomic field operators for the spin state $\left\vert
1,i\right\rangle $ as $\hat{\Psi}_{i}$ for species 1 and $\hat{\Phi}_{i}$
for species 2, the Hamiltonian for the mixture system in the second
quantization is represented by
\begin{eqnarray}
\hat{H} &=&\hat{H}_{1}+\hat{H}_{2}+\hat{H}_{12}, \\
\hat{H}_{1} &=&\int d\mathbf{r}\left\{ \hat{\Psi}_{i}^{\dag }(\frac{\hbar
^{2}}{2M_{1}}\nabla ^{2}+U_{1})\hat{\Psi}_{i}+\frac{\alpha _{1}}{2}\hat{\Psi}%
_{i}^{\dag }\hat{\Psi}_{j}^{\dag }\hat{\Psi}_{j}\hat{\Psi}_{i}\right.  \notag
\\
&&\left. +\frac{\beta _{1}}{2}\hat{\Psi}_{i}^{\dag }\hat{\Psi}_{j}^{\dag }%
\mathbf{F}_{1il}\cdot \mathbf{F}_{1jk}\hat{\Psi}_{k}\hat{\Psi}_{l}\right\} ,
\notag \\
\hat{H}_{12} &=&\frac{1}{2}\int d\mathbf{r}\left\{ \alpha \hat{\Psi}%
_{i}^{\dag }\hat{\Phi}_{j}^{\dag }\hat{\Phi}_{j}\hat{\Psi}_{i}\right.  \notag
\\
&&\left. +\beta \hat{\Psi}_{i}^{\dag }\hat{\Phi}_{j}^{\dag }\mathbf{F}%
_{1il}\cdot \mathbf{F}_{2jk}\hat{\Phi}_{k}\hat{\Psi}_{l}+\frac{\gamma }{3}%
\hat{O}^{\dag }\hat{O}\right\} .
\end{eqnarray}%
$H_{2}$ is the same as $H_{1}$ with the substitution of subscript $1$ by $2$
and $\hat{\Psi}_{i}$ by $\hat{\Phi}_{i}$ and $\hat{O}=\hat{\Psi}_{1}\hat{\Phi%
}_{-1}-\hat{\Psi}_{0}\hat{\Phi}_{0}+\hat{\Psi}_{-1}\hat{\Phi}_{1}$.

Through the control of the trapping frequency, we can make the two species
sufficiently overlapped and adopt the single spatial-mode approximation
(SMA) \cite{Law98,Pu99,Yi} for each of the two spinor condensates
with modes $\Psi (\mathbf{r})$ and $\Phi (\mathbf{r})$, i.e.
\begin{equation}
\hat{\Psi}_{i}=\hat{a}_{i}\Psi ,\text{\qquad }\hat{\Phi}_{i}=\hat{b}_{i}\Phi,
\end{equation}%
with $\hat{a}_{i}$ ($\hat{b}_{i}$) the annihilation operator for the
ferromagnetic (polar) atoms satisfying $\left[ \hat{a}_{i},\hat{a}_{j}\right]
=0$ and $\left[ \hat{a}_{i},\hat{a}_{j}^{\dag }\right] =\delta _{ij}$ (and
the same form of commutations for $\hat{b}_{i}$). The density-density
interaction part is a constant. Hence we only focus on the spin-dependent
Hamiltonian
\begin{eqnarray}
\hat{H} &=&\frac{c_{1}\beta _{1}}{2}(\mathbf{\hat{F}}_{1}^{2}-2\hat{N}_1)+%
\frac{c_{2}\beta _{2}}{2}(\mathbf{\hat{F}}_{2}^{2}-2\hat{N}_2)  \notag \\
&&+\frac{c_{12}\beta }{2}\mathbf{\hat{F}}_{1}\cdot \mathbf{\hat{F}}_{2}+%
\frac{c_{12}\gamma }{6}\hat{\Theta}_{12}^{\dag }\hat{\Theta}_{12},
\label{Ham}
\end{eqnarray}%
with $\mathbf{\hat{F}}_{1}=\hat{a}_{i}^{\dag }\mathbf{F}_{1ij}\hat{a}_{j}$ ($%
\mathbf{\hat{F}}_{2}=\hat{b}_{i}^{\dag }\mathbf{F}_{2ij}\hat{b}_{j}$)
defined in terms of the $3\times 3$ spin-1 matrices $\mathbf{F}_{1ij}$ $(%
\mathbf{F}_{2ij})$. The operator
\begin{equation}
\hat{\Theta}_{12}^{\dag }=\hat{a}_{0}^{\dag }\hat{b}_{0}^{\dag }-\hat{a}%
_{1}^{\dag }\hat{b}_{-1}^{\dag }-\hat{a}_{-1}^{\dag }\hat{b}_{1}^{\dag },
\end{equation}%
creates a singlet pair with one atom each from the two species, in much the
same way that the operators
\begin{equation}
\hat{A}^{\dag }=(\hat{a}_{0}^{\dag })^{2}-2\hat{a}_{1}^{\dag }\hat{a}%
_{-1}^{\dag },\text{\qquad }\hat{B}^{\dag }=(\hat{b}_{0}^{\dag })^{2}-2\hat{b%
}_{1}^{\dag }\hat{b}_{-1}^{\dag },
\end{equation}%
create pairs with two atoms from the same species \cite{HoYip,Koashi}. 
The interaction coefficients are $c_{1}=\int d\mathbf{r}\left\vert
\Psi (r)\right\vert ^{4},$ $c_{2}=\int d\mathbf{r}\left\vert \Phi
(r)\right\vert ^{4}$ and $c_{12}=\int d\mathbf{r}\left\vert \Psi
(r)\right\vert ^{2}\left\vert \Phi (r)\right\vert ^{2}$.

In the following we perform all analysis in the single-mode regime in the
absence of external fields. As learned from previous studies \cite{Yi},
SMA is shown to be exact for atomic interaction of the ferromagnetic type.
For polar interaction, if the magnetization $M=0$, the SMA wave function
is still exact and becomes invalid only if $M$ is large. In our mixture, we can 
safely apply the SMA to wave function of both polar and ferromagnetic
atoms in the absence of an external magnetic field. The ground state for polar atoms
is a fragile fragmented state with $M=0$. Small external fields or spatial 
dependence drive the system to symmetry broken states which are 
better captured by mean field theory. Very recently the symmetry broken
phase has been found in this binary mixture in the presence of a weak 
magnetic field \cite{XuBA}.

In our previous study \cite{zj} we report the anomalous fluctuations for the
numbers of atoms in the mixture of $^{23}$Na (polar) and $^{87}$Rb
(ferromagnetic) condensates in their $F=1$ manifold. DIA has been adopted to
ignore the $\gamma$ term. The ferromagnetic $^{87} $Rb condensate provides a
smooth background where the quantum many body states are hardly affected by
the fluctuation, while the fragile polar atoms $^{23}$Na are easier
influenced. Especially in the ground state of the AA phase, the interspecies
anti-ferromagnetic spin-exchange is large enough to polarize both species
and a maximally entangled state is realized between two species with total
spin $F=0$.

\section{The situation of $\protect\gamma =0$}

In this section, we consider the simple situation of $\gamma =0$ with
Hamiltonian
\begin{equation}
\hat{H}_{A}=\frac{c_{1}\beta _{1}}{2}\mathbf{\hat{F}}_{1}^{2}+\frac{%
c_{2}\beta _{2}}{2}\mathbf{\hat{F}}_{2}^{2}+\frac{c_{12}\beta }{2}\mathbf{%
\hat{F}}_{1}\cdot \mathbf{\hat{F}}_{2}.  \label{Ham2}
\end{equation}%
or in an alternative form
\begin{equation}
\hat{H}=a\mathbf{\hat{F}}_{1}^{2}+b\mathbf{\hat{F}}_{2}^{2}+c\mathbf{\hat{F}}%
^{2}  \label{Ham3}
\end{equation}%
where $a=c_{1}\beta _{1}/2-c_{12}\beta /4,b=c_{2}\beta _{2}/2-c_{12}\beta
/4,c=c_{12}\beta /4,$ and $\mathbf{\hat{F}}=\mathbf{\hat{F}}_{1}+\mathbf{%
\hat{F}}_{2}$ is the total angular momentum operator. A constant energy
shift of $c_j \beta _j N_j$ has been trivially eliminated. The results may
serve as reference states for the complete ground-state phases.

The eigenstates of the Hamiltonian (\ref{Ham3}) are then simply the common
eigenstates of the commutative operators $\mathbf{\hat{F}}_{1}^{2},\mathbf{%
\hat{F}}_{2}^{2},\mathbf{\hat{F}}^{2}$ and $\hat{F}_{z}$, denoted by
\begin{equation}
\left\vert F_{1},F_{2},F,m\right\rangle
=\sum_{m_{1}m_{2}}C_{F_{1,}m_{1};F_{2,}m_{2}}^{F,m}\left\vert
F_{1},m_{1}\right\rangle \left\vert F_{2},m_{2}\right\rangle
\end{equation}%
where the states in uncoupled representation
\begin{eqnarray*}
&&\left\vert F_{1},m_{1}\right\rangle =Z_{1}^{-\frac{1}{2}}(\hat{F}%
_{1-})^{F_{1}-m_{1}}(\hat{a}_{1}^{\dag })^{F_{1}}(\hat{A}^{\dag })^{\left(
N_{1}-F_{1}\right) /2}\left\vert 0\right\rangle \\
&&\left\vert F_{2},m_{2}\right\rangle =Z_{2}^{-\frac{1}{2}}(\hat{F}%
_{2-})^{F_{2}-m_{2}}(\hat{b}_{1}^{\dag })^{F_{2}}(\hat{B}^{\dag })^{\left(
N_{2}-F_{2}\right) /2}\left\vert 0\right\rangle
\end{eqnarray*}%
span a Hilbert space of dimension $(N_{1,2}+1)(N_{1,2}+2)/2$, respectively
\cite{Koashi}. Here $C$'s are the Clebsch-Gordon coefficients, $%
Z_{1,2}$ are the normalization constants and $\hat{F}_{1-}(\hat{F}_{2-})$ is
the lowering operator for $m_{1}$($m_{2}$). Minimizing the corresponding
eigenenergies of the Hamiltonian
\begin{equation}
E=aF_{1}(F_{1}+1)+bF_{2}(F_{2}+1)+cF(F+1),
\end{equation}%
we can get the ground state energy determined by different parameters $%
c_{1}\beta _{1}$, $c_{2}\beta _{2}$ and $c_{12}\beta $. In Fig. 1, using the
full quantum approach of exact diagonalization we calculate ground state
order parameters $\left\langle \mathbf{\hat{F}}_{1}^{2}\right\rangle ,$ $%
\left\langle \mathbf{\hat{F}}_{2}^{2}\right\rangle $ and $\left\langle
\mathbf{\hat{F}}_{1}\cdot \mathbf{\hat{F}}_{2}\right\rangle $ to illustrate
the different phases as $c_{12}\beta $ changes. The results are presented
for three typical cases (a) $c_{1}\beta _{1}=-1,c_{2}\beta _{2}=-2$, (b) $%
c_{1}\beta _{1}=1,c_{2}\beta _{2}=2$ and (c) $c_{1}\beta _{1}=-1,c_{2}\beta
_{2}=2$ with equal atomic numbers in two species, i.e. $N_{1}=N_{2}=N=100$.
We see that the results agree fairly well with the simulated annealing
approach in the mean field theory \cite{Xu09} except for some small
deviations originated from pure quantum effect. For instance,
the maximum and minimum values are $N(N+1)=10100$ and $0$ for $%
\left\langle \mathbf{\hat{F}}_{1}^{2}\right\rangle$ and $\left\langle
\mathbf{\hat{F}}_{2}^{2}\right\rangle $, but they are $N^2=10000$ and $%
-N(N+1)=-10100$ for $\left\langle \mathbf{\hat{F}}_{1}\cdot \mathbf{\hat{F}}%
_{2}\right\rangle $.

\begin{figure}[tbh]
\includegraphics[width=3.25in]{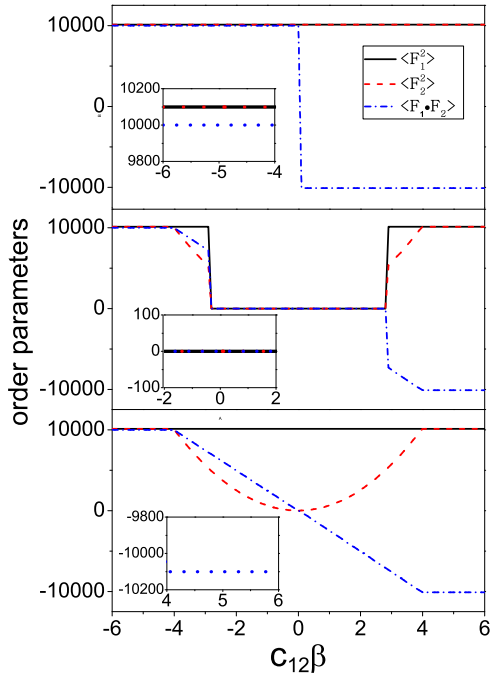}
\caption{(Color online) The dependence of ground-state order parameters on $%
c_{12}\protect\beta $ at fixed values of (a) $c_{1}\protect\beta _{1}=-1,$ $%
c_{2}\protect\beta _{2}=-2$; (b) $c_{1}\protect\beta _{1}=1,$ $c_{2}\protect%
\beta _{2}=2$; (c) $c_{1}\protect\beta _{1}=-1,$ $c_{2}\protect\beta _{2}=2$%
; and $c_{12}\protect\gamma = 0$ (all in units of $\left\vert c_{1}\protect%
\beta _{1}\right\vert $). Black solid lines, red dashed lines and blue
dot-dashed lines denote respectively the order parameters $\left\langle
\mathbf{\hat{F}}_{1}^{2}\right\rangle ,$ $\left\langle \mathbf{\hat{F}}%
_{2}^{2}\right\rangle $ and $\left\langle \mathbf{\hat{F}}_{1}\cdot \mathbf{%
\hat{F}}_{2}\right\rangle $. }
\label{fig1}
\end{figure}

\subsection{The case $c_{1}\protect\beta _{1}<0,$ $c_{2}\protect\beta _{2}<0$%
}

For the mixture of two ferromagnetic condensates, there are generally two
phases FF and AA separated by the critical point $0$ as shown in Fig. 1a.
The FF phase is described by a set of degenerate states generated by
repeatedly applying the lowering operators $(\hat{F}_{1-}+\hat{F}_{2-})$ on
the extreme states for $(F-m)$ times
\begin{equation}
\left\vert F_{1},F_{2},F,m\right\rangle =(\hat{F}_{1-}+\hat{F}%
_{2-})^{F-m}\left\vert F_{1},F_{2},F,F\right\rangle ,  \label{FF}
\end{equation}%
with $m =0,\pm 1,...\pm F$. The extreme states
\begin{equation}
\left\vert F_{1},F_{2},F,F\right\rangle =C_{N,N;N,N}^{2N,2N}\left\vert
N,N\right\rangle \left\vert N,N\right\rangle
\end{equation}%
can be simply described as $Z^{1/2}(\hat{a}_{1}^{\dag })^{N}(\hat{b}%
_{1}^{\dag })^{N}\left\vert 0\right\rangle .$ The AA phase is a singlet $%
\left\vert N,N,0,0\right\rangle $ with all states obey the condition $%
m_{1}+m_{2}=0$. All channels of total spin zero have to be taken into
account and we have%
\begin{equation}
\left\vert N,N,0,0\right\rangle =\underset{m_{1}=-N}{\overset{N}{\sum }}%
C_{N,m_{1};N,-m_{1}}^{0,0}\left\vert N,m_{1}\right\rangle \left\vert
N,-m_{1}\right\rangle .
\end{equation}
\ \

\subsection{The case $c_{1}\protect\beta _{1}>0,$ $c_{2}\protect\beta _{2}>0$%
}

The mixture of two polar condensates allows for five distinct phases FF, MM$%
_{-},$ PP, MM$_{+},$ and AA separated by four critical points $-(2N-1)
c_{2}\beta _{2}/N ,-c_{1}\beta _{1}-c_{2}\beta _{2},c_{1}\beta
_{1}+c_{2}\beta _{2}$ and $(2N-1) c_{2}\beta _{2}/(N+1)$ corresponding to $%
c_{12} \beta \simeq -4$, $-3$, $3,$ and $4$ in Fig. \ref{fig1}b.

In the region $c_{12}\beta \in (-(2N-1) c_{2}\beta _{2}/N ,-c_{1}\beta
_{1}-c_{2}\beta _{2})$, the MM$_{-}$ phase takes the same form as eq. (\ref%
{FF}), with extreme states represented by
\begin{equation}
\left\vert F_{1},F_{2},F,F\right\rangle
=C_{N,N;F_{2},F_{2}}^{N+F_{2},N+F_{2}}\left\vert N,N\right\rangle \left\vert
F_{2},F_{2}\right\rangle
\end{equation}%
or $Z^{1/2}(\hat{a}_{1}^{\dag })^{N}(\hat{b}_{1}^{\dag })^{F_{2}}(\hat{B}%
^{\dag })^{(N-F_{2})/2}\left\vert 0\right\rangle$. In MM$_{-}$ phase the
atoms in species 1 are totally polarized in one direction and form a
\textquotedblleft steady magnetic field\textquotedblright\ (black solid
line), and those in species 2 are partially polarized in the same direction.
Increasing the strength of coupling interaction ($\left\vert c_{12}\beta
\right\vert $) breaks singlet pairs in species 2 one by one, and results in
the increase of the total spin.

In the region $c_{12}\beta $ $\in (-c_{1}\beta _{1}-c_{2}\beta
_{2},c_{1}\beta _{1}+c_{2}\beta _{2})$, the two species are essentially
independent for a weak interspecies spin-exchange interaction. The PP phase
is a total spin singlet described by the direct product of the well known
polar ground state $Z^{-\frac{1}{2}}(\hat{A}^{\dag })^{N/2}(\hat{B}^{\dag
})^{N/2}\left\vert 0\right\rangle $ \cite{HoYip,Koashi}, giving
rise to $\left\langle \mathbf{\hat{F}}_{1}^{2}\right\rangle =0,$ $%
\left\langle \mathbf{\hat{F}}_{2}^{2}\right\rangle =0,$ and $\left\langle
\mathbf{\hat{F}}_{1}\cdot \mathbf{\hat{F}}_{2}\right\rangle=0$.

In the region of $c_{12}\beta \in (c_{1}\beta _{1}+c_{2}\beta
_{2},(2N-1)c_{2}\beta _{2}/(N+1))$, however, the MM$_{+}$ phase favors that
the atoms in species 2 are polarizing to the opposite direction of those in
species 1 and the total spin gradually decreases. This situation is much
more complicated because all states that satisfy the condition $%
m_{1}+m_{2}=N-F_{2}$ are involved and the summation index runs over all
possible Clebsch-Gordon coefficients, giving rise to the following extreme
state
\begin{equation}
\left\vert F_{1},F_{2},F,F\right\rangle =\underset{m_{1},m_{2}}{\sum }%
C_{N,m_{1};F_{2,}m_{2}}^{N-F_{2},N-F_{2}}\left\vert N,m_{1}\right\rangle
\left\vert F_{2},m_{2}\right\rangle
\end{equation}

\subsection{The case $c_{1}\protect\beta _{1}<0,$ $c_{2}\protect\beta _{2}>0$%
}

In the case of a mixture of a ferromagnetic and a polar condensate, four
possible phases FF, MM$_{-},$ MM$_{+},$ and AA are separated by three
critical points $-(2N-1)c_{2}\beta _{2}/N,0$ and $(2N-1) c_{2}\beta
_{2}/(N+1)$ corresponding to $c_{12} \beta =-4,0,4$ in Fig. 1 (c).

It has been shown that the ground state of polar atoms ($c_{2}\beta _{2}>0$)
only is a fragmented condensate with anomalously large number fluctuations
and fragile stability \cite{HoYip}, which can be described as $(\hat{B}%
^{\dag })^{N/2}\left\vert 0\right\rangle $. The $\hat{B}^{\dag }$ and $\hat{B%
}$ are invariant under any rotation of the system, and commute with $\hat{F}%
_{2}$ and $\hat{F}_{2z}$. However, for a ferromagnetic condensate ($%
c_{1}\beta _{1}<0$) the ground state favors that all atoms are aligned in
the same direction (i.e., $(\hat{a}_{1}^{\dag })^{N}\left\vert
0\right\rangle $) and much more stable. So when these two kinds of atoms are
mixed together, the polar atoms are more easier to be influenced, but their
back action on to the stable ferromagnetic atoms is negligible. This can be
seen from the constant black solid line in Fig.\ref{fig1} (c).

An interesting observation is that for large and negative (positive) value
of parameter $c_{12} \beta$ the system enters the same phase FF (AA), no
matter how the atoms interact inside each species. Detailed calculation show
that in FF phase all atoms are polarized in the same direction and the total
spin reaches its maximum value. The ground state is highly degenerate with
degeneracy $2F+1$. Take the state $\left\vert F_{1},F_{2},F,m=0\right\rangle
$ as an example, the atomic populations are
\begin{eqnarray}
\left\langle n_{0}^{(j)}\right\rangle&=&\frac{2N^2}{4N-1}  \notag \\
\left\langle n_{\pm 1}^{(j)}\right\rangle&=&\frac{N^2-N/2}{4N-1}
\end{eqnarray}
which reduces to $(N/4,N/2,N/4)$ for each species for large $N$. The number
fluctuations
\begin{eqnarray}
\left\langle \Delta n_{0}^{(j)}\right\rangle & =&\frac{2N}{4N-1}\sqrt {\frac{%
4N^{2}-9N/2+1}{4N-3}}  \notag \\
\left\langle \Delta n_{\pm 1}^{(j)}\right\rangle & =&\frac{N/2}{4N-1}\sqrt {%
\frac{32N^{2}-34N+7}{4N-3}}
\end{eqnarray}
are in order of $\sqrt{N}$ for large $N$.
On the other hand, in AA phase, atoms in species 1 and 2 are fully
polarized, however, in opposite directions and the total spin vanishes \cite%
{zj}. The singlet ground state is a fragmented condensate \cite{HoYip,zj} 
and the single particle density matrix is diagonal with atoms
equally populated $\left\langle n_{0,\pm 1}^{(j)}\right\rangle =N/3$ or $%
(N/3,N/3,N/3)$ for each species. The fluctuations
\begin{eqnarray}
\left\langle \Delta n_{0}^{(j)}\right\rangle &=&\frac{\sqrt{N^{2}+9N}}{3%
\sqrt{5}}  \notag \\
\left\langle \Delta n_{\pm 1}^{(j)}\right\rangle &=&\frac{2\sqrt{N^{2}+3N/2}%
}{3\sqrt{5}}.  \label{flu}
\end{eqnarray}
are anomalously large (in order of $N$) thus the fragmented condensate is
fragile.

\section{The situation of $\protect\gamma \neq 0$}

The last term $c_{12}\gamma $ in Hamiltonian (\ref{Ham}) is related to the
inter-species singlet paring. Although [$\mathbf{\hat{F}}^{2},\hat{\Theta}%
_{12}^{\dag }\hat{\Theta}_{12}]=0$, we notice that [$\mathbf{\hat{F}}%
_{1}^{2},\hat{\Theta}_{12}^{\dag }\hat{\Theta}_{12}]\neq 0$, and [$\mathbf{%
\hat{F}}_2^{2},\hat{\Theta}_{12}^{\dag }\hat{\Theta}_{12}]\neq 0$. Thus in
general they do not belong a set of commutative operators and the system is
not solvable. In the special case of $c_{1}\beta _{1}$=$c_{2}\beta _{2}$=$%
\frac{1}{2}c_{12}\beta $, we found that the spin-dependent Hamiltonian
reduces to a sum of commutative operators. The eigenstates can be
constructed by several building blocks \cite{XuZF, Koashi, HoYin,Ueda} 
which can be found via generating function method. To see more clearly
the role played by the $\gamma $ term, we focus on a special case in which $%
c_{12} \gamma$ is much larger than the other parts of the Hamiltonian and $%
c_{1}\beta _{1}<0$, $c_{2}\beta _{2}>0$. We know that the $\gamma $ term
encourages pairing two different types of atoms into singlets when $\gamma <0
$. In this case the interspecies singlet-pairing interaction dominates the
system and the total spin vanishes, namely $\left\langle \mathbf{\hat{F}}%
^{2}\right\rangle =\left\langle (\mathbf{\hat{F}}_{1}+\mathbf{\hat{F}}%
_{2})^{2}\right\rangle =0.$ There are two typical cases obeying the above
condition. In one case we have $\left\langle \mathbf{\hat{F}}%
_{1}^{2}\right\rangle =0,$ $\left\langle \mathbf{\hat{F}}_{2}^{2}\right%
\rangle =0,$ and $\left\langle \mathbf{\hat{F}}_{1}\cdot \mathbf{\hat{F}}%
_{2}\right\rangle=0$, which means the singlet formation occurs inside each
species, and the atoms in the same species are all paired with no net spin
left. In the other case we have $-2\left\langle \mathbf{\hat{F}}_{1}\cdot
\mathbf{\hat{F}}_{2}\right\rangle=\left\langle \mathbf{\hat{F}}%
_{1}^{2}\right\rangle +\left\langle \mathbf{\hat{F}}_{2}^{2}\right\rangle
\neq 0$, which means some intra-species pairs are broken, meanwhile singlet
formation occurs between different species. Under the condition $%
N_{1}=N_{2}=100$, numerical results show that the above two cases indeed
exist and well determined by the parameter $c_{12}\beta$.
\begin{figure}[t]
\includegraphics[width=3.25in]{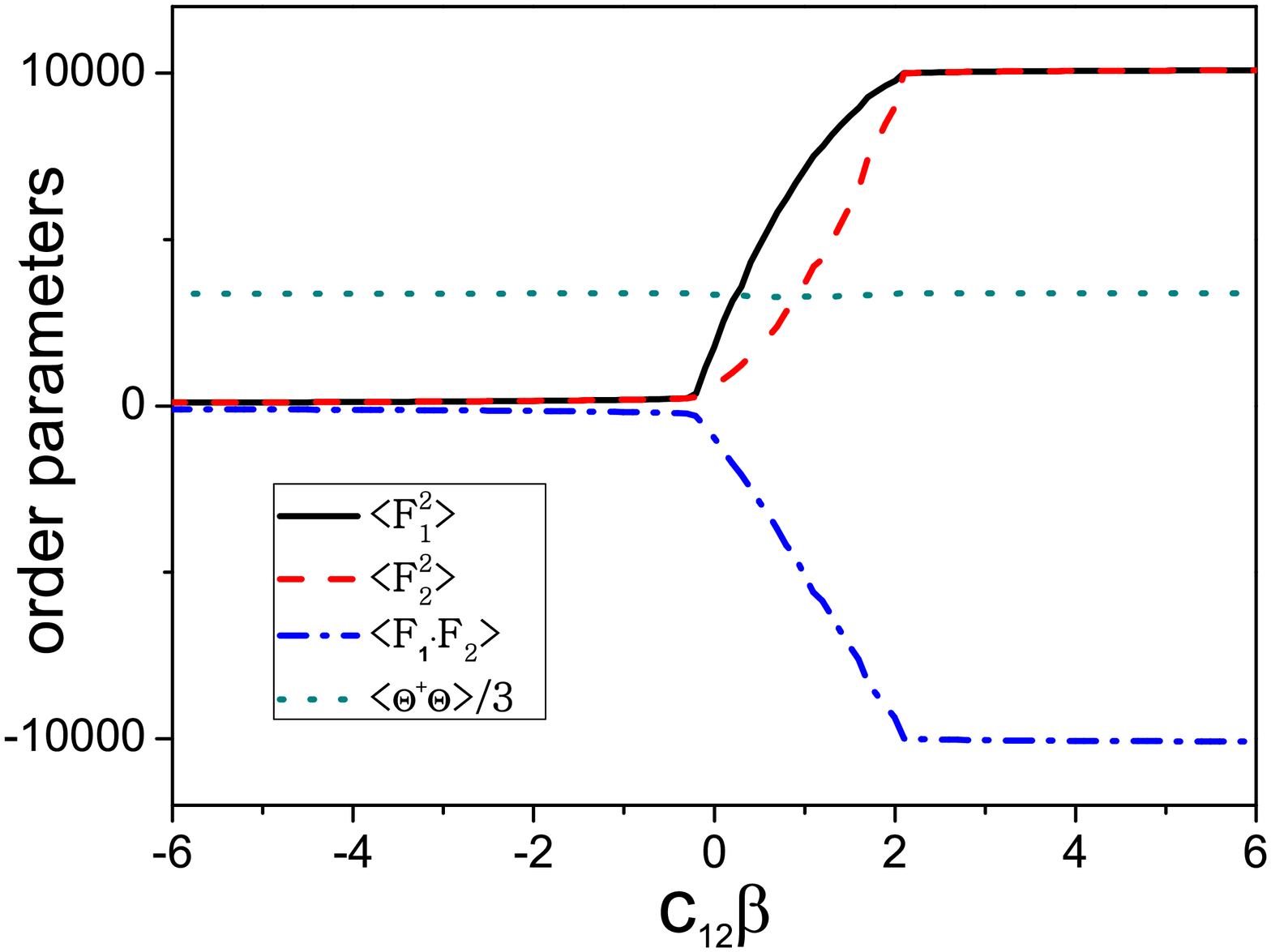}
\caption{(Color online) The dependence of ground-state order parameters on $%
c_{12}\protect\beta $ at fixed values of $c_{1}\protect\beta _{1}=-1,$ $c_{2}%
\protect\beta _{2}=2,$ and $c_{12}\protect\gamma =-20$ (in the unit of $%
\left\vert c_{1}\protect\beta _{1}\right\vert $). Black solid lines, red
dashed lines, blue dot-dashed lines and green dotted lines denote
respectively the order parameters $\left\langle \mathbf{\hat{F}}%
_{1}^{2}\right\rangle ,$ $\left\langle \mathbf{\hat{F}}_{2}^{2}\right\rangle
$, $\left\langle \mathbf{\hat{F}}_{1}\cdot \mathbf{\hat{F}}_{2}\right\rangle$%
, and $\left\langle \hat{\Theta }_{12}^{\dag }\hat{\Theta }%
_{12}\right\rangle /3.$}
\end{figure}

We consider the direct product of the Fock states of the two species $%
\left\vert n_{1}^{(1)},n_{0}^{(1)},n_{-1}^{(1)}\right\rangle \otimes
\left\vert n_{1}^{(2)},n_{0}^{(2)},n_{-1}^{(2)}\right\rangle $, which may be
equivalently defined as
\begin{eqnarray}
&&\hat{n}_{\alpha }^{(1,2)}\left\vert
n_{0}^{(1)},m_{1},n_{0}^{(2)},m_{2};m\right\rangle  \notag \\
&=&n_{\alpha }^{(1,2)}\left\vert
n_{0}^{(1)},m_{1},n_{0}^{(2)},m_{2};m\right\rangle .
\end{eqnarray}%
Here $m_{1,2}$ are the corresponding magnetization specified as $%
m_{1,2}=n_{1}^{(1,2)}-n_{-1}^{(1,2)}$ and $m=m_{1}+m_{2}$ is the total
magnetization. For simplification, we restrict ourselves into the subspace
that the total magnetization is conserved $m=0$, in which case all states
are non-degenerate. The Hamiltonian (\ref{Ham}) is then represented in a
sparse matrix and with the approach of exact diagonalization we numerically
get the ground state of the system, on which the order parameters are
calculated.
\begin{figure}[tb]
\includegraphics[width=3.5in]{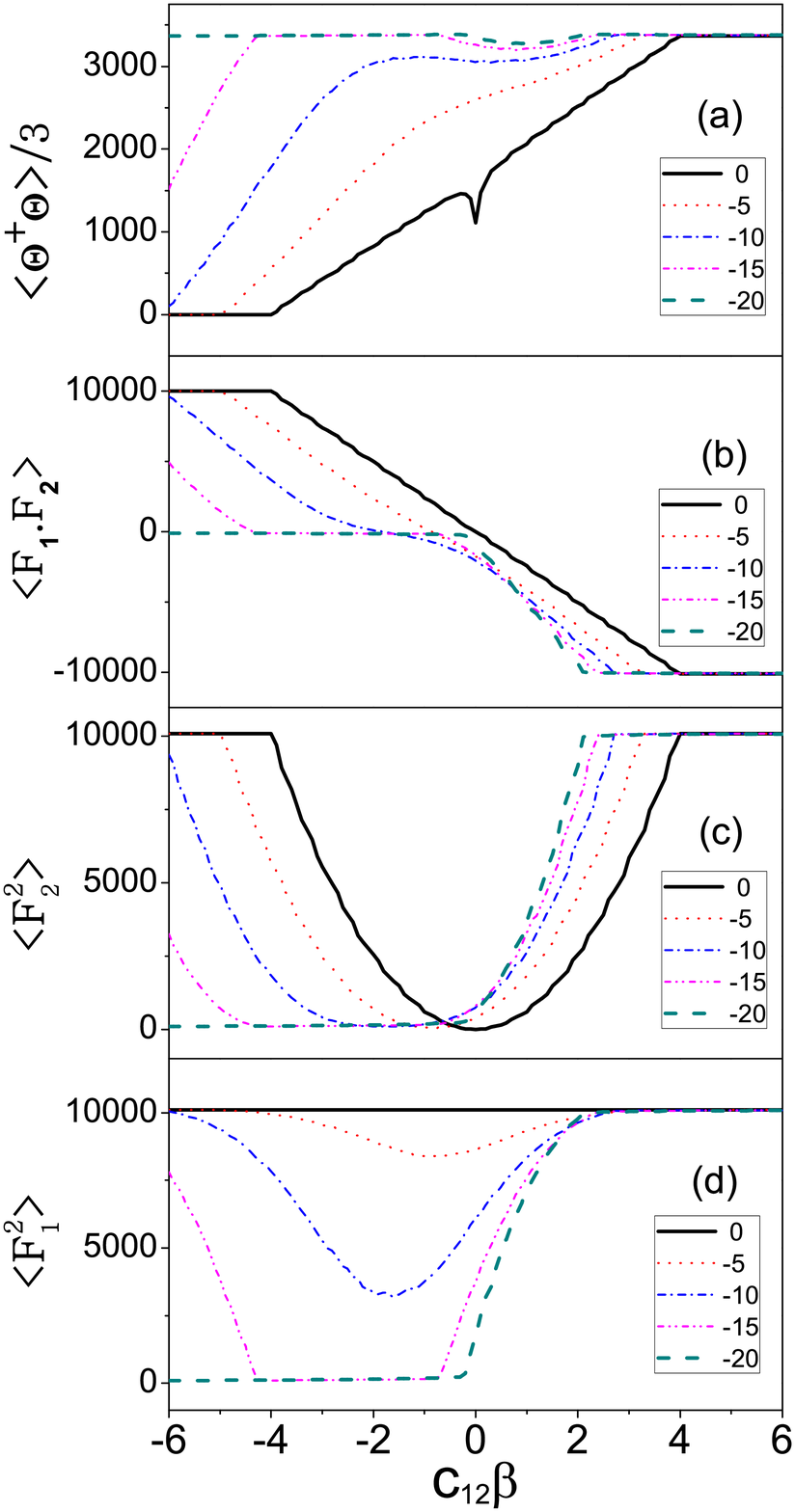}
\caption{(Color online) The dependence of the four order parameters
on the parameter $c_{12} \beta$ for different inter-species paring interaction parameter $c_{12}\protect\gamma= 0, -5,
-10, -15, -20$. }
\end{figure}
Fig. 2 shows the dependence of the four order parameters on $c_{12} \beta$
at fixed values of $c_{1}\beta _{1}=-1,$ $c_{2}\beta _{2}=2,$ and $%
c_{12}\gamma =-20$. We find that in the region of $c_{12}\beta <0$, $%
\left\langle \mathbf{\hat{F}}_{1}^{2}\right\rangle$, $\left\langle \mathbf{%
\hat{F}}_{2}^{2}\right\rangle $, and $\left\langle \mathbf{\hat{F}}_{1}\cdot
\mathbf{\hat{F}}_{2}\right\rangle$ are all approximatively equal to zero,
corresponding to the first case of vanishing total spin. However, for
positive $c_{12}\beta$, atoms in each species begin to polarize and are
fully polarized for $c_{12}\beta \geq 2$, with the negative value of $%
\left\langle \mathbf{\hat{F}}_{1}\cdot \mathbf{\hat{F}}_{2}\right\rangle$
showing that they are polarized to the opposite directions. During the whole
process the three order parameters always obey the condition $-2\left\langle
\mathbf{\hat{F}}_{1}\cdot \mathbf{\hat{F}}_{2}\right\rangle=\left\langle
\mathbf{\hat{F}}_{1}^{2}\right\rangle +\left\langle \mathbf{\hat{F}}%
_{2}^{2}\right\rangle $. The distinction in the envelope of $\left\langle
\mathbf{\hat{F}}_{1}^{2}\right\rangle $ and $\left\langle \mathbf{\hat{F}}%
_{2}^{2}\right\rangle $ in the region $0<c_{12}\beta <\,2$ lies in the
choice of $c_{1}\beta _{1}=-1,$ $c_{2}\beta _{2}=2$.

The fourth order parameter $\left\langle \hat{\Theta}_{12}^{\dag }\hat{\Theta%
}_{12}\right\rangle /3$ reflects the feature of the total spin $F$ of the
system. It takes a constant value, which is actually the maximum value, in
entire region of $c_{12}\beta $ in Figure 2 (see the green line), while the
total spin reaches its minimum value $0$. In order to see more clearly the
competition between the interspecies coupling and singlet pairing
interactions, in Figure 3 we illustrate four order parameters for different
values of $c_{12}\gamma $. An obvious variation can be easily seen from, for example,  Figure 3d,
where the black solid line ($\gamma=0$) gradually changes to the purple dashed line ($\gamma =-20$).
We find that when $\gamma $ =0 the
order parameters $\left\langle \hat{\Theta}_{12}^{\dag }\hat{\Theta}%
_{12}\right\rangle /3$ becomes zero in the FF phase implying that the total spin $F$
gets to its maximum (Fig. 3a), while in the AA phase it equals to a constant value and 
the total spin $F$ vanishes. In between the two limits \textquotedblleft
0\textquotedblright\ and "constant", the intermediate value of the order parameters $\left\langle \hat{%
\Theta}_{12}^{\dag }\hat{\Theta}_{12}\right\rangle /3$ indicates that the
system is a mixture with both singlet pairs and nonzero net magnetization. This
feature can be well understood for the special case when $\gamma =0$
and $\beta =0$, marked as a small dip in the black solid line in Fig. 3a. We find that $%
\left\langle \hat{\Theta}_{12}^{\dag }\hat{\Theta}_{12}\right\rangle /3=10000/9$, which can be 
obtained analytically. 
In fact, the state at this point can be simply expressed as a
direct product of two well-known states (ferromagnetic and polar), $(%
\hat{a}_{1}^{\dag })^{N}\left\vert 0\right\rangle \otimes (\hat{B}%
^{\dag})^{N/2}\left\vert 0\right\rangle$, and we find that on this state
\begin{eqnarray*}
\frac{1}{3}\left\langle \hat{\Theta}_{12}^{\dag }\hat{\Theta}%
_{12}\right\rangle  &=&\frac{1}{3}\left\langle \hat{a}_{1}^{\dag }%
\hat{a}_{1}{}\hat{b}_{-1}^{\dag }{}{}\hat{b}_{-1}{}-\hat{a}%
_{0}^{\dag }\hat{a}_{1}{}\hat{b}_{0}^{\dag }{}{}\hat{b}_{-1}{}+%
\hat{a}_{-1}^{\dag }\hat{a}_{1}{}\hat{b}_{1}^{\dag }{}{}\hat{%
b}_{-1}\right. {} \\
&&-\hat{a}_{1}^{\dag }\hat{a}_{0}{}\hat{b}_{-1}^{\dag }{}{}%
\hat{b}_{0}{}+\hat{a}_{0}^{\dag }\hat{a}_{0}{}\hat{b}%
_{0}^{\dag }{}{}\hat{b}_{0}{}-\hat{a}_{-1}^{\dag }\hat{a}_{0}{}%
\hat{b}_{1}^{\dag }{}{}\hat{b}_{0} \\
&&\left. +\hat{a}_{1}^{\dag }\hat{a}_{-1}{}\hat{b}_{-1}^{\dag
}{}{}\hat{b}_{1}{}-\hat{a}_{0}^{\dag }\hat{a}_{-1}{}\hat{b}%
_{0}^{\dag }{}{}\hat{b}_{1}{}+\hat{a}_{-1}^{\dag }\hat{a}_{-1}{}%
\hat{b}_{1}^{\dag }{}{}\hat{b}_{1}\right\rangle  \\
&=&\frac{1}{3}\left\langle \hat{a}_{1}^{\dag }\hat{a}_{1}{}\hat{b%
}_{-1}^{\dag }{}{}\hat{b}_{-1}{}\right\rangle  \\
&=&\frac{1}{3}\cdot N\cdot \frac{N}{3}=N^{2}/9 \label{3states}
\end{eqnarray*}%
which agrees with the numerical result. We notice that the system is a
mixture with $N/2$ polar pairs and net magnetization  $F=F_{1}=N$.

In our system there exist three states, on which the average of total spin amounts to zero. 
Although we have $\mathbf{\hat{F}}^{2}(\hat{A}^{\dag})^{N/2}({\hat{B}}%
^{\dag})^{N/2}\left\vert 0\right\rangle =0,$ $\mathbf{\hat{F}}^{2}\left\vert
N,N,0,0\right\rangle =0,$ and $\mathbf{\hat{F}}^{2}(\hat{\Theta}_{12}^{\dag
})^{N}\left\vert 0\right\rangle =0$, the difference between these states can be easily seen from an 
example $N_{1}=N_{2}=N=2$. We find that
\begin{eqnarray}
\left\vert 2,2,0,0\right\rangle  &=&Z^{1/2}\underset{m_{1},m_{2}}{\sum }%
C_{F_{1,}m_{1};F_{2,}m_{2}}^{F=0,m=0}\left\vert 2,m_{1}\right\rangle
\left\vert 2,m_{2}\right\rangle  \notag \\
&=&\frac{1}{48}\{C_{2,0;2,0}^{0,0}(\hat{F}_{1-})^{2-0}(\hat{F}_{2-})^{2-0}(\hat{a}%
_{1}^{\dag })^{2}(\hat{b}_{1}^{\dag })^{2}\left\vert 0\right\rangle    \notag \\
&&\text{+}C_{2,1;2,-1}^{0,0}(\hat{F}_{1-})^{2-1}(\hat{F}_{2-})^{2+1}(\hat{a}%
_{1}^{\dag })^{2}(\hat{b}_{1}^{\dag })^{2}\left\vert 0\right\rangle   \notag  \\
&&\text{+}C_{2,2;2,-2}^{0,0}(\hat{F}_{1-})^{2-2}(\hat{F}_{2-})^{2+2}(\hat{a}%
_{1}^{\dag })^{2}(\hat{b}_{1}^{\dag })^{2}\left\vert 0\right\rangle    \notag \\
&&\text{+}C_{2,-1;2,1}^{0,0}(\hat{F}_{1-})^{2+1}(\hat{F}_{2-})^{2-1}(\hat{a}%
_{1}^{\dag })^{2}(\hat{b}_{1}^{\dag })^{2}\left\vert 0\right\rangle    \notag \\
&&\text{+}C_{2,-2;2,2}^{0,0}(\hat{F}_{1-})^{2+2}(\hat{F}_{2-})^{2-2}(\hat{a}%
_{1}^{\dag })^{2}(\hat{b}_{1}^{\dag })^{2}\left\vert 0\right\rangle \}   \notag \\
&=&\frac{1}{2\sqrt{5}}((\hat{\Theta}_{12}^{\dag })^{2}-\frac{1}{3}\hat{A}^{\dag}%
\hat{B}^{\dag})\left\vert 0\right\rangle 
\end{eqnarray}%
From the relation between these three states (\ref{3states}), we see that the AA phase of our system 
includes at lest two pairing mechanism, i.e. $\hat{\Theta}%
_{12}^{\dag }$ and $\hat{A}^{\dag}\hat{B}^{\dag}$.
As total spin $F$ vanishes, the number distributions of these three states are all $\left\langle n_{1}^{(j)}\right\rangle
=\left\langle n_{0}^{(j)}\right\rangle =\left\langle
n_{-1}^{(j)}\right\rangle =N/3$, but the number fluctuation on
these states are quite different. For the state $(%
\hat{A}^{\dag})^{N/2}(\hat{B}^{\dag})^{N/2}\left\vert 0\right\rangle $ \cite%
{HoYip}, the results are
\begin{eqnarray}
\left\langle \Delta n_{1}^{(j)}\right\rangle  &=&\left\langle \Delta
n_{0}^{(j)}\right\rangle /2=\left\langle \Delta n_{-1}^{(j)}\right\rangle  \notag \\
&=&\frac{\sqrt{N^{2}+3N}}{3\sqrt{5}}.
\end{eqnarray}%
On the state $\left\vert N,N,0,0\right\rangle $ \cite{zj}, we have obtained in eq. (\ref{flu}), in the case of large $N$, satisfying
\begin{equation}
\left\langle \Delta n_{1}^{(j)}\right\rangle  =2\left\langle \Delta
n_{0}^{(j)}\right\rangle =\left\langle \Delta n_{-1}^{(j)}\right\rangle,
\end{equation}%
while for the state $(\hat{\Theta}_{12}^{\dag })^{N}\left\vert
0\right\rangle ,$ we find that the number fluctuations are equally
distributed \cite{zj}, i.e.
\begin{eqnarray}
\left\langle \Delta n_{1}^{(j)}\right\rangle  &=&\left\langle \Delta
n_{0}^{(j)}\right\rangle =\left\langle \Delta n_{-1}^{(j)}\right\rangle
\notag \\
&=&\sqrt{\frac{N(N+1)}{6}-\frac{N^{2}}{9}}.
\end{eqnarray}%

\begin{figure}[tbph]
\includegraphics[width=3.0in]{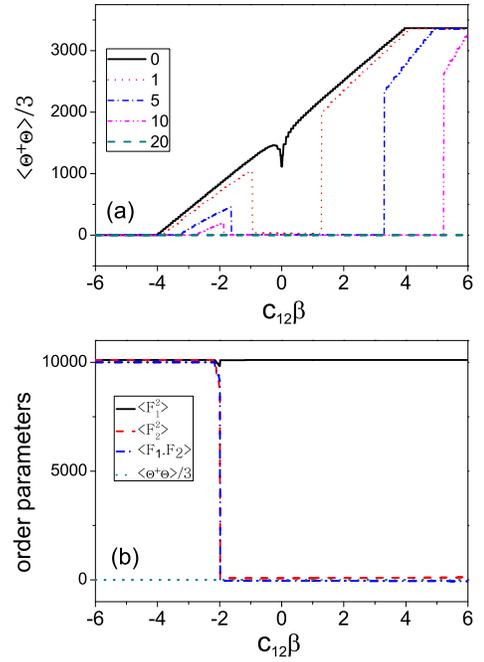}
\caption{(Color online) (a) The dependence of the fourth order parameter
on the parameter $c_{12} \beta$ for different inter-species paring interaction parameter $c_{12}\protect\gamma= 0, 1, 5,
10, 20$. (b) The dependence of ground-state order parameters on $c_{12}%
\protect\beta $ at fixed values of $c_{1}\protect\beta _{1}=-1,$ $c_{2}%
\protect\beta _{2}=2,$ and $c_{12}\protect\gamma =20.$ }
\label{1}
\end{figure}

Finally we discuss briefly the situation when $c_{12}\gamma >0$, in which case 
the ground state favors minimizing the $\gamma $ term. 
We illustrate the competition between the interspecies coupling and singlet pairing 
interactions numerically in Fig. 4a. If $c_{12}\gamma $ is far more larger than any other
parameters, the order parameter $\left\langle \hat{\Theta }_{12}^{\dag }\hat{\Theta }%
_{12}\right\rangle /3$ will vanish (see the dashed line in Figure 4a). 
All order parameters take the value
$0$ or $\simeq N^{2}$ when $c_12 \gamma \gg 1$, with the boundary determined by the corresponding
amplitudes of spin-coupling strengths as shown in Fig. 4b for $c_{12}  \gamma =20$.

\section{Conclusion}

In summary we studied the interspecies singlet pairing in the ground state 
of a binary mixture of spin-1 condensates in the absence of a magnetic 
field. In the case of $c_{12} \gamma =0$,
the exact quantum states can be constructed from angular momentum theory
for the mixture of two ferromagnetic, two polar, and ferromagnetic-polar condensates. The
ground state is classified into five types according to the inter-species
coupling parameter $c_{12}\beta $. By means of the full quantum approach of exact
diagonalization, more general case of $\gamma \neq 0$ is considered. We
illustrate the competition between the two interspecies interaction $%
c_{12}\beta $ and $c_{12}\gamma$, and find that if $c_{12}\gamma \ll -1$
the ground state is a singlet of the total spin. There, however, exist different
types of singlet formations determined by $c_{12}\beta$. 

This work is supported by the NSF of China under Grant Nos. 10774095 and 11074153, the National Basic Research
Program of China (973 Program) under Grant Nos. 2010CB923103, 2011CB921601, the NSF
of Shanxi Province, Shanxi Scholarship Council of China, and the Program for New Century Excellent
Talents in University (NCET).


\begin{thebibliography}{99}
\bibitem{Stamper-Kurn1} D. M. Stamper-Kurn, M. R. Andrews, A. P.
Chikkatur, S. Inouye, H.-J. Miesner, J. Stenger, and W. Ketterle, Phys. Rev.
Lett. \textbf{80}, 2027 (1998).

\bibitem{UedaRev} M. Ueda and Y. Kawaguchi, arXiv:1001.2072

\bibitem{Stamper-Kurn2} H.-J. Miesner, D. M. Stamper-Kurn, J. Stenger,
S. Inouye, A. P. Chikkatur, and W. Ketterle, Phys. Rev. Lett. \textbf{82},
2228 (1999).

\bibitem{Ohmi} T. Ohmi and K. Machida, J. Phys. Soc. Jpn. \textbf{67}, 1822
(1998).


\bibitem{Law98} C. K. Law, H. Pu, and N. P. Bigelow, Phys. Rev. Lett.
\textbf{81}, 5257 (1998).

\bibitem{Pu99} H. Pu, C. K. Law, S. Raghavan, J. H. Eberly, and N. P.
Bigelow, Phys. Rev. A \textbf{60}, 1463 (1999).

\bibitem{WenxianZhang} W. Zhang, D. L. Zhou, M-S. Chang, M. S. Chapman, and
L. You, Phys. Rev. A \textbf{72}, 013602 (2005)

\bibitem{HoYip} T.-L. Ho and S.-K. Yip, Phys. Rev. Lett. \textbf{84},
4031 (2000).


\bibitem{Ho} T.-L. Ho, Phys. Rev. Lett. \textbf{81}, 742 (1998).

\bibitem{Mueller06} E. J. Mueller, T.-L. Ho, M. Ueda, and G. Baym, Phys.
Rev. A \textbf{74}, 033612 (2006).

\bibitem{Stenger} J. Stenger, S. Inouye, D. M. Stamper-Kurn, H.-J. Miesner, A. P. Chikkatur, W. Ketterle, Nature (London) \textbf{396}, 345
(1998).

\bibitem{MSChang} M.-S. Chang, C. D. Hamley, M. D. Barrett, J. A. Sauer,
K.M. Fortier, W. Zhang, L. You, and M. S. Chapman, Phys. Rev. Lett. \textbf{%
92}, 140403 (2004).

\bibitem{Koashi} M. Koashi and M. Ueda, Phys. Rev. Lett. \textbf{84},
1066 (2000).

\bibitem{Ueda} M. Ueda and M. Koashi, Phys. Rev. A \textbf{65}, 063602
(2002).

\bibitem{Ciobanu} C. V. Ciobanu, S.-K. Yip, and T.-L. Ho, Phys. Rev. A
\textbf{61}, 033607 (2000).

\bibitem{Ho96} T.-L. Ho and V. B. Shenoy, Phys. Rev. Lett. \textbf{77}, 3276
(1996).

\bibitem{Pu2} H. Pu and N. P. Bigelow, Phys. Rev. Lett. \textbf{80}, 1130
(1998); ibid, \textbf{80}, 1134 (1998)

\bibitem{Esry} B. D. Esry, C.H. Greene, J. P. Burke, Jr., and J. L. Bohn,
Phys. Rev. Lett. \textbf{78}, 3594 (1997).

\bibitem{Timmermans} E. Timmermans, Phys. Rev. Lett. \textbf{81}, 5718
(1998).

\bibitem{Myatt} C. J. Myatt, E. A. Burt, R. W. Ghrist, E. A. Cornell, and C.
E. Wieman, Phys. Rev. Lett. \textbf{78}, 586 (1997).

\bibitem{Modugno} G. Modugno, M. Modugno, F. Riboli, G. Roati, and M.
Inguscio, Phys. Rev. Lett. \textbf{89}, 190404 (2002).

\bibitem{Thalhammer} G. Thalhammer, G. Barontini, L. De Sarlo, J. Catani, F.
Minardi, and M. Inguscio, Phys. Rev. Lett. \textbf{100}, 210402 (2008).

\bibitem{Papp} S. B. Papp, J. M. Pino, and C. E. Wieman, Phys. Rev. Lett.
\textbf{101}, 040402 (2008).

\bibitem{OFR} P. O. Fedichev, Yu. Kagan, G. V. Shlyapnikov, and J. T. M. Walraven, 
Phys. Rev. Lett. \textbf{77}, 2913 (1996); F. K. Fatemi, K. M. Jones, and P. D. Lett, \textit{ibid.} 
\textbf{85}, 4462 (2000); M. Theis, G. Thalhammer, K. Winkler, M. Hellwig,
G. Ruff, R. Grimm, and J. H. Denschlag, \textit{ibid.} \textbf{93, }123001
(2004); C. D. Hamley, E. M. Bookjans, G. Behin-Aein, P. Ahmadi,
and M. S. Chapman,  Phys. Rev. A \textbf{79}, 023401 (2009).

\bibitem{Xu09} Z. F. Xu, Y. Zhang, and L.You, Phys. Rev. A \textbf{79},
023613 (2009).

\bibitem{XuZF} Z. F. Xu, J. Zhang, Y. Zhang, and L. You, Phys. Rev. A
\textbf{81}, 033603 (2010).

\bibitem{zj} J. Zhang, Z. F. Xu, L. You and Y. Zhang, Phys. Rev. A \textbf{82%
}, 013625 (2010).

\bibitem{XuBA} Z. F. Xu, J. W. Mei, R. L\"{u}, and L.You, Phys. Rev. A \textbf{82}, 053626 (2010) .

\bibitem{Luo} M. Luo, Z. Li, and C. Bao, Phys. Rev. A \textbf{75}, 043609
(2007).

\bibitem{Zhang2010} W. Zhang, B. Sun, M. S. Chapman, and L. You, Phys. Rev. A
\textbf{81}, 033602 (2010).

\bibitem{DIA} A. C. Maan, E. Tiesinga, H. T. C. Stoof, and B. J. Verhaar,
Physica. B. \textbf{165}, 17 (1990).

\bibitem{DIA1} S. B. Weiss, M. Bhattacharya, and N. P. Bigelow, Phys. Rev. A
\textbf{68}, 042708 (2003); \textit{ibid.} \textbf{69}, 049903(E) (2004).

\bibitem{DIA2} A. Pashov, O. Docenko, M. Tamanis, R. Ferber, H. Knoeckel, and
E. Tiemann, Phys. Rev. A 72, 062505 (2005).

\bibitem{Yi} S. Yi, \"{O}. E. Mustecaplioglu, C. P. Sun, and L. You,
Phys. Rev. A \textbf{66}, 011601(R) (2002).



\bibitem{HoYin} T.-L. Ho and L. Yin, Phys. Rev. Lett. \textbf{84}, 2302
(2000).
\end{thebibliography}
\end{document}